\documentclass[journal=jpccck]{achemso}
\usepackage{epsfig}
\usepackage{amsmath}
\setkeys{acs}{keywords = true}

\usepackage{subfigure}
\usepackage{graphicx}
\usepackage{dcolumn}
\usepackage{bm}
\usepackage[latin1]{inputenc}
\title{Strain-Tunable Spin Moment in Ni-Doped Graphene}

\author{Elton~J.~G.~Santos}  
\affiliation{Centro de F\'{\i}sica de Materiales (CFM-MPC),
Centro Mixto CSIC-UPV/EHU, 
Paseo Manuel de Lardizabal 5, 20018 San Sebasti\'an, Spain}
\alsoaffiliation{Donostia International Physics Center (DIPC),
Paseo Manuel de Lardizabal 4, 20018 San Sebasti\'an, Spain}

\author{A. Ayuela} 
\affiliation{Centro de F\'{\i}sica de Materiales (CFM-MPC),
Centro Mixto CSIC-UPV/EHU, 
Paseo Manuel de Lardizabal 5, 20018 San Sebasti\'an, Spain}
\alsoaffiliation{Donostia International Physics Center (DIPC),
Paseo Manuel de Lardizabal 4, 20018 San Sebasti\'an, Spain}

\author{D. S\'anchez-Portal} \email{sqbsapod@ehu.es}
\affiliation{Centro de F\'{\i}sica de Materiales (CFM-MPC),
Centro Mixto CSIC-UPV/EHU, 
Paseo Manuel de Lardizabal 5, 20018 San Sebasti\'an, Spain}
\alsoaffiliation{Donostia International Physics Center (DIPC),
Paseo Manuel de Lardizabal 4, 20018 San Sebasti\'an, Spain}


\date{\today}

\begin{document} 

\begin{abstract}
Graphene, due to its exceptional properties, 
is a promising material for nanotechnology applications.
In this context,
the ability to tune 
the properties of graphene-based materials and devices 
with the incorporation of defects and impurities can be of 
extraordinary importance. Here 
we investigate the effect of uniaxial tensile strain 
on the electronic and magnetic properties of graphene doped with 
substitutional Ni impurities (Ni$_{sub}$). We have found that,   
although Ni$_{sub}$ defects are non-magnetic in the relaxed
layer, uniaxial
strain induces a spin moment in the system.
The spin moment
increases with the applied strain up to values of 0.3-0.4$\mu_B$
per Ni$_{sub}$, until a critical strain of $\sim$6.5\% is reached.
At this point, a sharp transition to a high-spin state ($\sim$1.9 $\mu_B$)
is observed.
This magnetoelastic effect could be utilized to design
strain-tunable spin devices based on Ni-doped graphene.

\end{abstract}

\section{Introduction}

The special electronic and magnetic properties
of graphene
have led to the proposal of
new types of electronic and spintronic
devices based 
on this material~\cite{Novoselov04,Novoselov05,Louie06,Geim07,Louie08,graph-review,Geim09_review}.
For this purpose, it is instrumental
to study the impact of defects and impurities on such properties.
Defects can deteriorate the performance
of graphene-based devices. However, defects and impurities can also be used
to tune the electronic, magnetic and mechanical 
properties of those devices and, therefore, be intentionally incorporated
in the structure. 
This has driven an increasing attention to 
defects in graphene~\cite{Banhart11}. 
Both intrinsic defects~\cite{Amara07,Lehtinen03}
and extrinsic defects, like 
substitutional atoms~\cite{Santos08,Co-paper10,NJP10,Arkadi09},
are presently under intense research.
In this context, substitutional metal impurities seem
particularly attractive. Recent experiments
indicate that metal atoms can be easily incoporated in the 
graphenic layer as
substitutional dopants at predefined positions.
Carbon vacancies can be created in graphene and carbon
nanotubes by focusing an electron beam using scanning 
transmission electron microscope. Metal atoms migrating
on the graphenic surfaces are then observed to be trapped in such
defects with bonding energies of several eVs~\cite{Manzo2010,Cretu2010}.
This opens a route to fabricate stable arrays of subtitutional
impurities that may allow for the exploration 
of the interesting magnetic and electronic
properties predicted for such systems~\cite{Santos08,Co-paper10}.

Ni substitutional impurities (Ni$_{sub}$) have been experimentally
detected in single-walled carbon nanotubes (SWCNTs)~\cite{Ushiro06}
and large fullerenes and graphitic particles~\cite{Banhart00}.
Theoretical studies on the magnetic properties
of these defects revealed surprising results.
While Ni$_{sub}$ impurities are non-magnetic in
flat graphene~\cite{NJP10,Arkadi09}, they can develop a substantial
spin-moment (up to $\sim$0.8~$\mu_B$) in metallic
SWCNTs~\cite{Santos08}.
This peculiar behavior stems from the curvature dependence of the
electronic structure of Ni$_{sub}$ defects.
Therefore, in Ref.~\cite{Santos08} it was proposed that
curvature could be used to switch on the magnetism of
Ni$_{sub}$ impurities
in metallic graphenic nanostructures.
Unfortunately,
the local curvature at the location of the Ni$_{sub}$ defect
cannot be easily determined and controlled in experiments.

For this reason, in the present paper we explore a different
way to modify the local geometry and, consequently, the
electronic and magnetic properties of Ni$_{sub}$ impurities.
Here, we study the effect of
uniaxial tensile strain on graphene layers doped with Ni$_{sub}$ defects.
As we will see in detail below, we have found that uniaxial
strain can be used to induce a transition of the Ni$_{sub}$ defect
to a magnetic state and, thus, to tune the magnetic
properties of Ni-doped graphene. 
At moderate
strains, Ni$_{sub}$ defects develop a spin moment 
that slowly increases with strain up to values of 0.3-0.4$\mu_B$.
The change is more dramatic when a critical 
strain of $\sim$6.5\% is reached:
An abrupt increase of the spin moment to $\sim$1.9 $\mu_B$
is observed.
This magnetoelastic effect could be utilized to design 
strain-tunable spin devices based on Ni-doped graphene. 

Substitutional transition metal impurities in graphene 
and, in general, small clusters of 
transition metals interacting with graphenic 
layers are likely sites for the 
adsorption of molecules~\cite{Lastra10,Yoo09}. 
For example, the possible catalytic activity of Au atoms 
embedded in graphene for CO oxidation has been theoretically 
studied~\cite{Lu09}. Experiments have also found an enhanced 
catalytic activity of this oxidation reaction for sub-nanometer Pt clusters 
deposited on graphene~\cite{Yoo09}. 
There are also recent calculations of oxygen interacting 
with Cr and Mn substitutionals in graphene~\cite{Dai10}. 
The spin moment of these impurities change after the 
adsorption of the molecule and this possibility should be taken into account 
when comparing the present results with measurements under ambient 
conditions. An additional complication is that the reactivity of the 
impurities is likely to depend on the applied strain~\cite{Zhou10,Zhou10bis}. 
Therefore, we restrict our study to the evolution of the spin moment for
substitutionally Ni-doped graphene under vacuum conditions 
as a function of a well-defined 
external parameter, the applied uniaxial strain.

\section{Methods}

To theoretically investigate the problem
we have carried out first-principles density-functional 
calculations of Ni$_{sub}$ defects
in graphene 
under uniaxial strains in an 
experimentally accessible range\cite{Geim09,Kim09}. 
The calculations have been performed using 
the {\sc Siesta} code~\cite{Soler02} and
the PBE-GGA functional~\cite{pbe-functional}.
We have used a rectangular supercell containing 
111 C atoms and a single Ni$_{sub}$ impurity. 
The carbon layers are always separated by a distance 
of at least 17~\AA, so that the interaction between adjacent layers is negligible.
Stresses of different magnitudes are applied along 
the $(n,n)$ and $(n,0)$ directions, and 
both atomic coordinates and lattice 
vectors are allowed to relax
\bibnote{
From our calculations we can extract the
Young modulus (E)  and the Poisson ratio ($\nu$) of the system.
For pristine graphene we obtain E=57~eV/atom and $\nu$=0.15
in good agreement with other calculations and the
experimental data~\cite{PRB99}. The Ni$_{sub}$ impurity
in our supercell has a small impact on these parameters,
we obtain E$\sim$55 eV/atom and $\nu\sim$0.17 for the Ni-doped
layer. }.
A double-$\zeta$ polarized (DZP)~\cite{Soler02}
basis set has been always used for the calculation of the magnetic
and electronic properties. 
However, we have checked that using a
double-$\zeta$ (DZ) basis set yields almost identical
relaxed structures and spin moments as the DZP basis and, therefore,
we have used the smaller DZ basis for most 
structural relaxations. 
Other computational parameters are similar to those
used in our previous work on Ni$_{sub}$
impurities in SWCNTs~\cite{Santos08}
and graphene~\cite{NJP10}.
We have also performed GGA+U calculations at different values
of the strain using the
formulation of Dudarev {\it et al.}~\cite{Dudarev98} 
and several values of the U parameter 
varied in the range 1-4~eV.
GGA+U results did not show significant deviations from 
PBE-GGA calculations and, thus, here we only present the
latter.

\section{Results and Discussion}

\begin{figure}
\includegraphics[width=4.0000in]{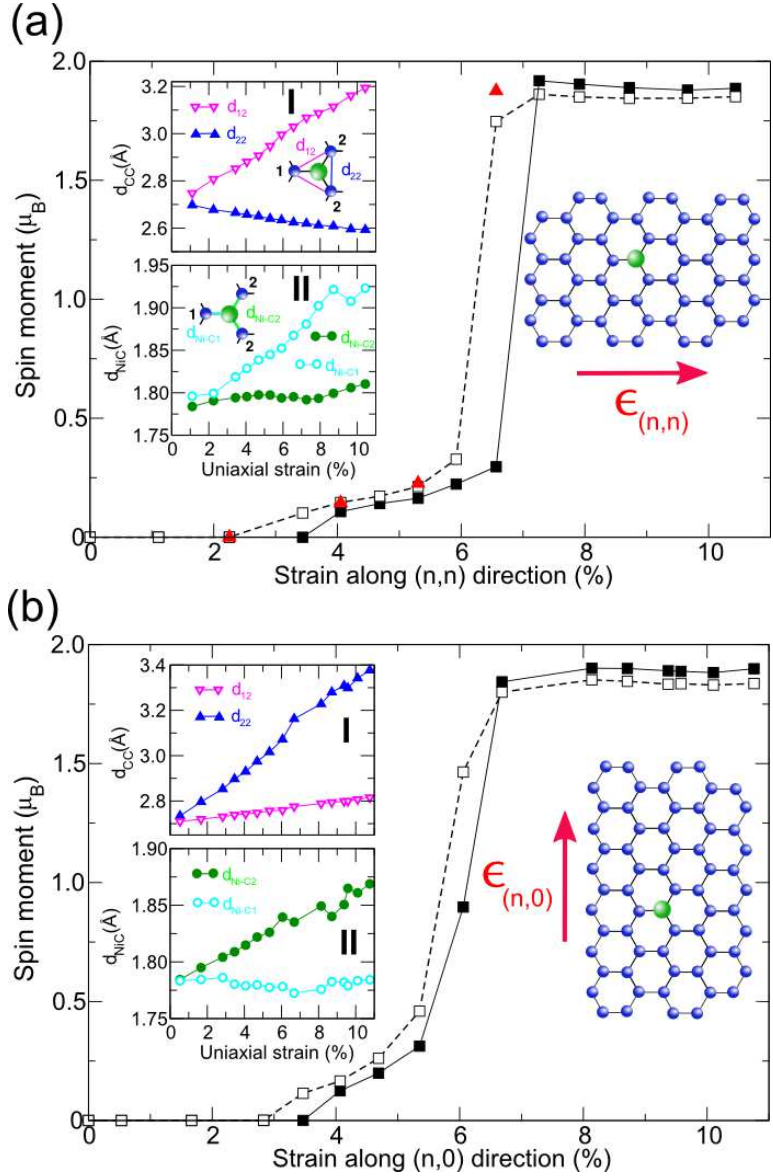}
\caption{(Color online) Spin moment of a Ni$_{sub}$ impurity 
in graphene as a function of the applied strain 
($\epsilon$) along the $(n,n)$ (a) and 
$(n,0)$ (b) directions. A schematic illustration of the 
structure is included. In panels (a) and (b), 
filled squares indicate results 
obtained using geometries from a non-spin polarized calculation 
using a DZ basis. The 
electronic structure and spin moment are calculated 
using a more complete DZP basis using such geometry.
Open squares indicate a similar calculation, 
but the geometries come from a spin-polarized 
calculation in this case. The triangles represent 
a calculation where both the relaxed geometry and the spin moment have been
obtained using a DZP basis set. 
Insets I and II in panels (a) and (b) show the 
distances between carbon atoms C1 and C2 (d$_{12,22}$) and Ni-C bond lenghts 
(d$_{\text{Ni-C1, Ni-C2}}$) as a function of the applied 
uniaxial strain calculated using a DZ basis set and spin polarization.
}
\label{fig1}
\end{figure}

\ref{fig1} shows the spin moment of a Ni$_{sub}$ defect 
as a function of the strain applied along 
the $(n,n)$ (a) and $(n,0)$ (b) directions.
Open and filled squares show results obtained
using geometries relaxed
with and without spin polarization, respectively.
These relaxations utilized a DZ basis set, although the spin moments
in \ref{fig1} were
always computed using the more complete DZP basis.
The triangles display calculations where the
geometry was obtained using 
a DZP basis set and spin polarization. 
The relaxed structures and the behavior of the spin moment 
as a function of the uniaxial strain are 
very similar in all the cases.
At zero strain the Ni$_{sub}$ defect is non-magnetic as was previously 
reported and analyzed in earlier studies\cite{Santos08,NJP10}. 
The structure around the defect deforms as the uniaxial tension is applied
(see the insets
in \ref{fig1}). As a consequence, the electronic structure of the defect
is modified and a spin moment develops.  
We see that at $\sim$3.5\% strain the 
system becomes magnetic with a spin moment
that increases slowly
with the uniaxial strain up to values of 
$\sim$0.30-0.40~$\mu_{B}$ for a strain of $\sim$6.0\%.  
At $\sim$6.5\% strain, the spin moment 
increases sharply to $\sim$1.9~$\mu_{B}$
and remains almost constant for larger strains. 
This means that a moderate variation of the strain applied
on the layer can induce a large
change of the spin moment associated with the Ni$_{sub}$ defects.
This is particularly attractive in connection with recent
experimental reports that indicate that
metal atoms can 
be easily incorporated 
in graphene as substitutional dopants, even
at predefined position~\cite{Manzo2010,Cretu2010}.
Therefore, the present results can, in principle, be used to design a new
family of magnetoelastic devices based on graphene.
Furthermore, if uniaxial strain is varied in a larger
scale similar to the one presented in \ref{fig1}, 
the magnetism of Ni-doped graphene can be switched 
on and off at will. 
Since according to recent experiments~\cite{Geim09,Kim09} 
uniaxial strain can be applied on graphene in 
a controlled way, 
our results are suitable for experimental verification.

The transition to a high-spin solution in \ref{fig1} 
is similar for both strain orientations, although 
somewhat more abrupt 
for the $(n,n)$ direction (panel (a)) 
where no intermediate steps are observed. 
This small difference already points to the role played 
on the development of the spin moment 
by the local defect geometry and its 
orientation relative to the applied strain.
Insets I and II in \ref{fig1}(a) and (b) present  
this geometrical information. 
When the strain is applied, the largest deformation 
corresponds to the triangle formed 
by the three C neighbors of the Ni$_{sub}$ impurity. 
As expected, the variation of the 
three C-C distances (d$_{\text{C-C}}$) is different depending
on their orientation 
relative to the strain direction. The length of 
those C-C bonds pointing along the strain direction increases, 
whereas those pointing along the perpendicular direction 
are reduced by a much smaller amount. This corresponds to the 
expected elastic behavior for graphene,
and can be understood as a 
tendency to keep approximately a constant area per atom.
As we will see
below, this distortion strongly affects the position
of the defect levels associated with the Ni$_{sub}$ impurity.
The Ni-C bonds (d$_{\text{Ni-C}}$)
also increase on average. However, for the studied range of 
strains up to $\sim$10.5\%, the maximum increase of 
d$_{\text{Ni-C}}$ is $\sim$8\% whereas it 
is as large as $\sim$25\% for d$_{\text{C-C}}$.
This difference is partially explained by the elevated
position of the Ni atom over the graphene layer, that provides 
an additional degree of freedom to respond to the layer elongation. 
The Ni atom is larger than the C atom that was removed to create 
the vacancy, thus it moves out of the carbon plane.
At zero strain its height over the carbon layer is $\sim$0.9~\AA. 
This height decreases as a function of the applied strain 
(by as a much as 17\% for the maximum strain applied here). 
In spite of its elevated position, the Ni atom is strongly bound 
to the C vacancy with a large binding energy of about 7~eV~\cite{NJP10}, 
which clearly ensures its chemical stability.

\begin{figure}
\includegraphics[width=4.000in]{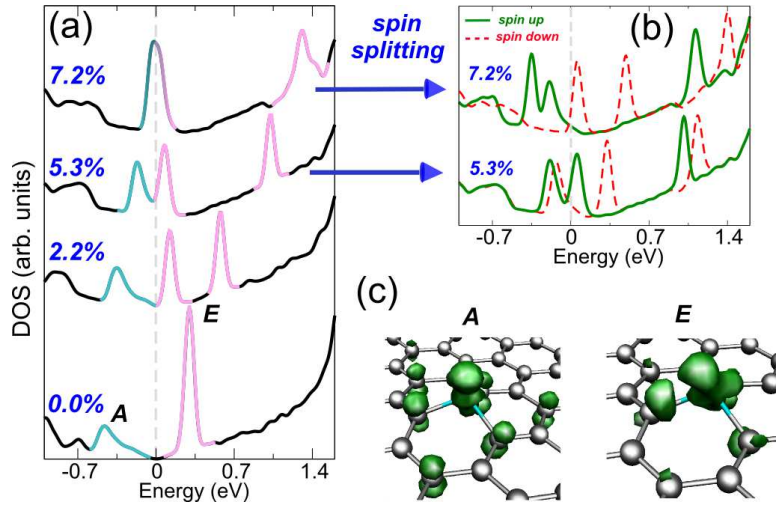}
\caption{(Color online) (a) Spin-unpolarized density of 
states (DOS) of a Ni$_{sub}$ impurity 
in graphene under uniaxial strains of 
0.0\%, 2.2\%, 5.3\% and 7.2\% along the $(n,n)$ 
direction. Symbols $A$ and $E$ indicate the symmetry of 
the defect levels coming from the hybridization of Ni 
and the neighboring C atoms at zero strain.
The level $A$ corresponds to a Ni $3d_{z^2}-$C $2p_{z}$ hybridization and 
the levels $E$ have a Ni $3d_{xz}, 3d_{yz}-$C $2sp$ character. 
(b) Spin-polarized DOS showing the spin splitting and the occupations
of the different levels at strains of 5.3\% and 7.2\%. 
Solid (green)
and dashed (red) lines stand 
for majority and minority spins, respectively.  
Different curves in panels (a) and (b) have been shifted for clarity.
The Fermi energy is marked by the dashed vertical line and is set to zero. 
(c) Density associated with  
the $A$ and the lowest $E$ levels at a strain of 5.3\% 
with an isovalue of $\pm$0.002 e$^-$/Bohr$^3$ (spin-unpolarized calculation).
}
\label{fig2-dos}
\end{figure}

In order to understand the origin of the
tunable spin-moment of Ni$_{sub}$ defects in graphene,
the densities of states (DOS) for spin-unpolarized 
calculations under 0.0\%, 2.2\%, 5.3\% and 7.2\% strain 
along the $(n,n)$ direction
are shown in \ref{fig2-dos}~(a). 
The qualitative 
behavior is similar if the strain is applied along 
other directions. At zero strain 
we find 
two sharp peaks around the Fermi energy (E$_F$).
They correspond to a singly- 
and a doubly-degenerate defect level, respectively.
The character
of these levels was analyzed in detail in Ref.~\cite{NJP10}.
They mainly come from the hybridization of the Ni 3$d$
orbitals with the neighboring C atoms. 
Due to the symmetric position of 
the metal atom over the vacancy, the system has a $C_{3v}$ 
symmetry at zero strain and the electronic levels can 
be classified according 
to the $A$ or $E$ irreducible representations of this point group. 
The position of these defect states as a function 
of the applied strain determines the observed behavior of
the spin moment.

One of the levels, with $A$ character, is 
occupied and appears around $\sim$0.5~eV 
below E$_F$ at zero strain. This state comes from a fully symmetric 
linear combination of the $2p_z$ orbitals (z-axis normal to the layer) 
of the  nearest C neighbors interacting with the $3d_{z^2}$ orbital
of Ni. The other twofold-degenerate level, with $E$ character, 
comes from the hybridization 
of the in-plane $sp$ lobes of the carbon neighbors with the 
Ni $3d_{xz}$ and $3d_{yz}$ orbitals and appears at 0.4~eV above E$_F$ 
at zero strain. 
As a consequence of this electronic structure, with 
the Ni $3d$ states far from E$_F$
and no flat bands crossing E$_F$, the magnetic 
moment of the Ni$_{sub}$ impurity in graphene is zero. 
Interestingly, these levels appearing close
to E$_F$ in \ref{fig2-dos}(a) are reminiscent of those found 
for the unreconstructed carbon vacancy in graphene\cite{NJP10,Amara07}.

The energy position of these three levels 
shifts as a function of the applied strain. This can be clearly seen in
\ref{fig2-dos}~(a). When a finite strain is applied, the degeneracy 
between Ni 3$d_{xz}-$C 2$sp$ and Ni 3$d_{yz}-$C 2$sp$ states ($E$ levels) 
is removed and one of them gradually shifts towards E$_{F}$. 
The fully-occupied Ni 3$d_{z^2}-$C 2$p_z$ state (with $A$ symmetry) 
also moves towards E$_F$ as a function of the strain, although 
at a somewhat smaller pace. For small strains (above $\sim$3.5\%) the lowest $E$ 
level becomes partially populated. At somewhat larger strains ($\sim$5\%), the $A$ 
level is also sufficiently close to E$_F$ so it starts to lose a small
part of its charge. 
This population is transferred to the lowest $E$ level, so the occupations of these two 
levels change without an appreciable modification of the total charge localized 
around the Ni impurity.  Finally, around 6.5\% strain both, the $A$ state and 
the lowest $E$ level, become half-occupied. This exchange of population between the $A$
and lowest $E$ levels is the mechanism behind the sharp transition of 
the spin moment observed in the calculations. Eventually both levels 
become half-occupied, so they polarize and the system develops a spin 
moment of 2~$\mu_B$.

This can be seen more clearly in \ref{fig2-dos}~(b), which presents 
the spin-polarized DOS for two strains around the transition point. 
Closely below the transition, at 5.3\% strain, the lowest $E$ level is 
partially populated, developing a small spin polarization. Simultaneously, 
the fully-occupied $A$ level starts to lose part of its population, 
although presents an almost negligible spin polarization. 
The corresponding exchange splittings are $\sim$0.29~eV and $\sim$0.13~eV 
for the $E$ levels and only $\sim$0.05~eV for the $A$ state. 
The larger splitting and corresponding polarization of the $E$ levels is
related to their larger localization. The spatial distribution of the density 
associated with the $A$ level and the lowest $E$ level can be seen in \ref{fig2-dos}~(c). 
The different localization and symmetry of the two states, as well as the 
antibonding nature of the interaction between Ni and the neighboring C 
atoms, is manifest in these plots. 
Above the transition, for example at 
7.2\% strain in \ref{fig2-dos}~(c), we can see that both levels, $A$ and the lowest $E$, 
are fully polarized. Accordingly, the spin splittings increase 
considerably: $\sim$0.86~eV and $\sim$0.34eV for the $E$ states 
and $\sim$0.25~eV for the A level. The stability of the 
spin-polarized solutions with respect to the non-magnetic 
solutions is also enhanced from 14 meV per Ni atom 
at 5.3\% strain to 184 meV at 7.2\% strain.

\begin{figure}
\includegraphics[width=4.000in]{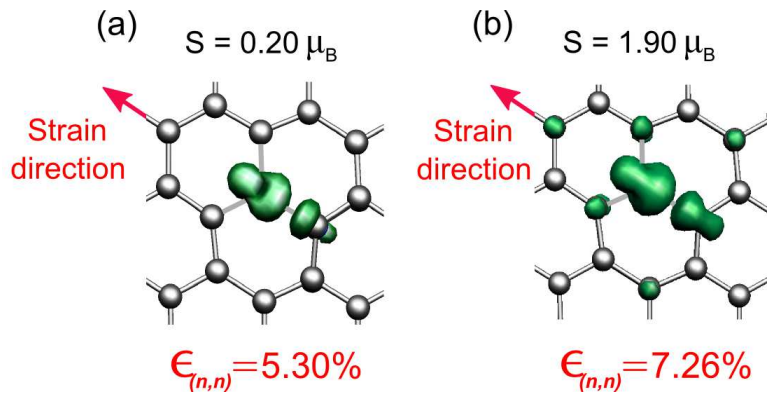}
\caption{(Color online) (a)-(b) Spin densities for Ni$_{sub}$ defects in 
graphene at strains of 5.30\% and 7.2\% along the $(n,n)$ direction. 
The strain direction
is marked by the red arrows in both panels. The isovalue cutoff 
in (a) and (b) panels 
is $\pm$0.035 and $\pm$0.060 e$^-$/Bohr$^3$, respectively. 
}
\label{fig3}
\end{figure}

The changes of the spin-polarization when moving across the transition
at $\sim$6.5\% strain are depicted in \ref{fig3}. 
The spin polarization
in the neighboring carbon atoms depends on the
strain intensity. At 5.30\% strain, the spin density is mainly
localized at the Ni impurity and at the C atom directly bonded
to it along the strain direction.
The magnetization density reflects the anti-bonding  
character of this $E$ defect state, as seen in the node along the bond.
The $2sp$-like shape of the spin density around the 
C atom is also clear. This should be contrasted with the spin
distribution at 7.2\% strain. At this larger
value of the strain,
the spin density presents an additional $2p_z$ component on the
neighboring carbon atoms. This substantial modification 
of the spin density
confirms the contribution from
the Ni $3d_{z^2}-$C $2p_{z}$ defect level for strains 
above $\sim$6.5\%, as already expected 
from the analysis presented in \ref{fig2-dos}.

According to our interpretation,
the moderate polarization 
of the Ni$_{sub}$ defect in flat graphene
as a function of the applied uniaxial strain below $\sim$6.5\% strain
is clearly reminiscent of 
the curvature-induced spin-polarization found in Ni$_{sub}$-doped
SWCNTs~\cite{Santos08}. In that case, it was shown that 
the curvature of the carbon wall in SWCNTs was able to break the degeneracy 
of the two unoccupied $E$-symmetry defect levels of the 
Ni$_{sub}$ impurities.
For sufficiently large
curvatures, in the case of metallic tubes, 
one of these levels would become partially populated 
and induce the spin polarization of the Ni$_{sub}$-SWCNT system.
However, there are a number of differences between the present
case and that of SWCNTs. One of them is that, due to the reduced
density of states near E$_{F}$ in graphene, the spin-polarization 
is more localized in the case of flat graphene under uniaxial strain. 
Another important difference is that the 
spin moment of Ni$_{sub}$ in graphene can, at least in principle, 
be easily tuned since the 
degree of strain applied
to the layer can be controlled~\cite{Geim09,Kim09}.
Finally, in the
case of graphene under tensile strain, for strains above 
a $\sim$6.5\% 
the Ni$_{sub}$ defects present a high-spin state that was
not found in the case of the curvature-induced magnetism.

\section{Conclusions}

In summary, our first-principles calculations show that
the electronic and magnetic structure of substitutional
Ni defects in graphene can be tuned with the use of uniaxial strain. 
While Ni$_{sub}$ defects are predicted to be non-magnetic 
for the unstrained layer, we observe that by stretching 
the layer by a few percents
it is possible to turn their magnetism on. 
The spin moment increases slowly with 
the applied strain until a critical value ($\sim$6.5\%) is reached.
At this value, the spin moment exhibits a sharp transition
to a higher value 
of $\sim$1.9~$\mu_B$ and remains almost constant as the
strain is futher increased. This result is consistent
with a very recent report by Huang {\it et al.}~\cite{Huang11} using
a different computational methodology. According to our calculations,
this behavior of the spin moment
is weakly 
dependent on the orientation of the applied strain. A detailed 
analysis indicates that this strain-tunable spin moment
results from changes in the position of three defect levels around 
the Fermi energy which are antibonding 
combinations of the Ni $3d$ states and the $2p_{z}$ 
and $2sp$ orbitals of the neighboring C atoms. 
This tunable magnetism observed in Ni$_{sub}$ defects 
may be interesting for spintronic devices based on graphene.

\section{Acknowledgements}

We acknowledge support from Basque
Departamento de Educacion and the UPV/EHU (Grant
No. IT-366-07), the Spanish Ministerio de Educaci\'on
y Ciencia (Grant No. FIS2010-19609-CO2-02) 
and the ETORTEK program funded by the Basque Departamento de 
Industria and the Diputacion Foral de Guipuzcoa.

\providecommand*\mcitethebibliography{\thebibliography}
\csname @ifundefined\endcsname{endmcitethebibliography}
  {\let\endmcitethebibliography\endthebibliography}{}

\newpage

\begin{figure}
\includegraphics[width=6.0000in]{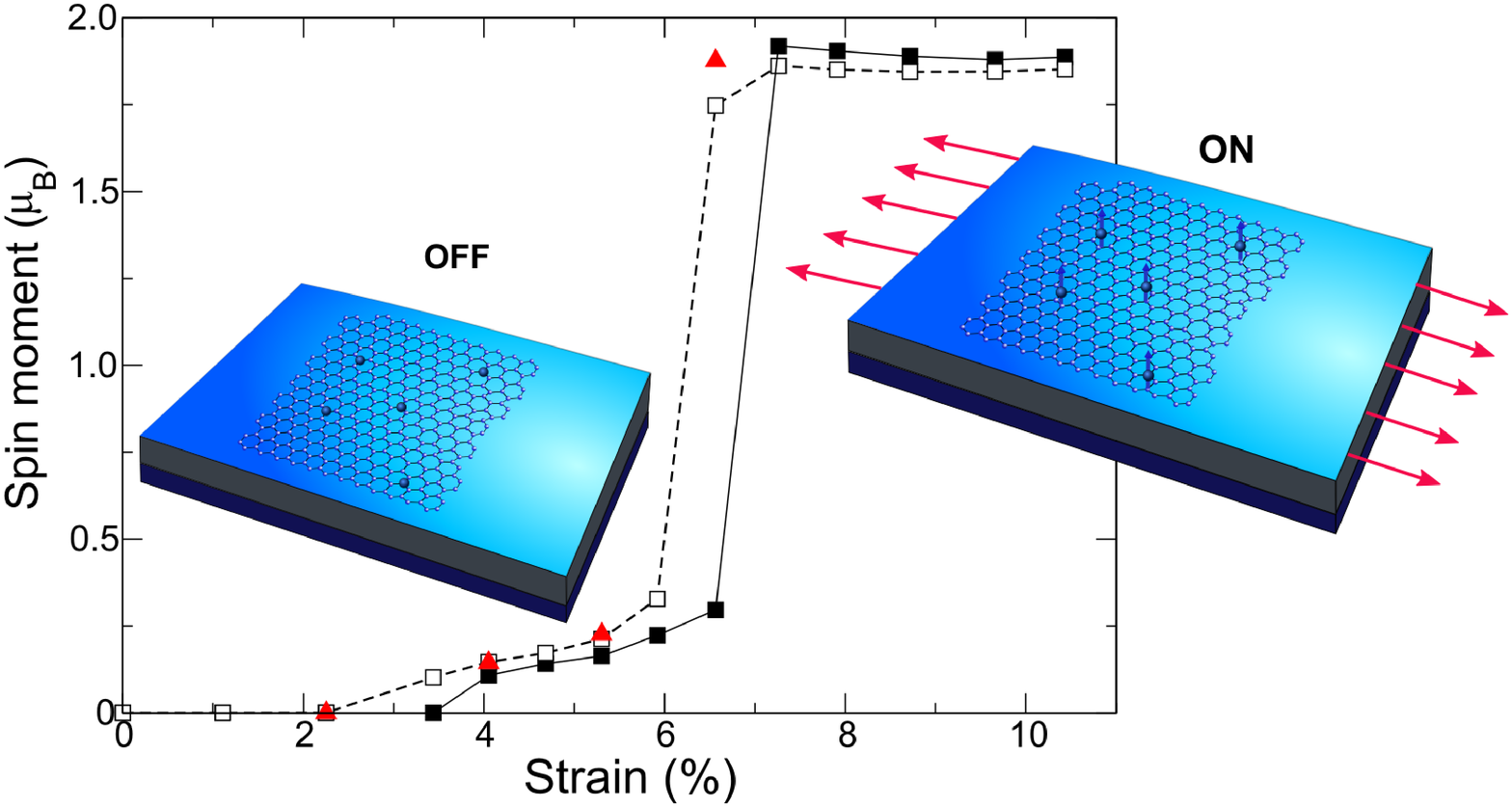}
\caption{ TOC }
\end{figure}


\begin{mcitethebibliography}{33}
\bibitem[Not()]{Note-1}
From our calculations we can extract the Young modulus (E) and the Poisson
  ratio ($\nu$) of the system. For pristine graphene we obtain E=57~eV/atom and
  $\nu$=0.15 in good agreement with other calculations and the experimental
  data~\cite{PRB99}. The Ni$_{sub}$ impurity in our supercell has a small
  impact on these parameters, we obtain E$\sim$55 eV/atom and $\nu\sim$0.17 for
  the Ni-doped layer.\relax
\mciteBstWouldAddEndPunctfalse
\mciteSetBstMidEndSepPunct{\mcitedefaultmidpunct}
{}{\mcitedefaultseppunct}\relax
\EndOfBibitem
\end{mcitethebibliography}
\end{document}